\RequirePackage{fix-cm}
\documentclass[smallextended]{svjour3}       
\smartqed  
\usepackage{graphicx}
\usepackage{amsmath}
\usepackage{amssymb}

\begin{document}

\title{Analytical Alignment Tolerances for Off-Plane Reflection Grating Spectroscopy
}

\titlerunning{Analytical OXRGS Alignment Tolerances}        

\author{Ryan Allured         \and
        Randall T. McEntaffer
}


\institute{Ryan Allured \at
Harvard-Smithsonian Center for Astrophysics, Cambridge, MA, USA \\
\email{rallured@cfa.harvard.edu} \\
\and
Randall T. McEntaffer \at
              University of Iowa, Iowa City, IA, USA \\
}

\date{Received: date / Accepted: date}

\maketitle

\begin{abstract}
Future NASA X-ray Observatories will shed light on a variety of high-energy astrophysical phenomena.  Off-plane reflection gratings can be used to provide high throughput and spectral resolution in the 0.3--1.5 keV band, allowing for unprecedented diagnostics of energetic astrophysical processes.  A grating spectrometer consists of multiple aligned gratings intersecting the converging beam of a Wolter-I telescope.  Each grating will be aligned such that the diffracted spectra overlap at the focal plane.  Misalignments will degrade both spectral resolution and effective area.  In this paper we present an analytical formulation of alignment tolerances that define grating orientations in all six degrees of freedom.  We verify our analytical results with raytrace simulations to fully explore the alignment parameter space.  We also investigate the effect of misalignments on diffraction efficiency.
\end{abstract}

\section{Introduction}

The development of critical technologies is required to accomplish the science goals of future NASA X-ray observatories.  One such technology is off-plane reflection gratings to produce high throughput and high spectral resolving power at energies below 1.5 keV.  Grating spectrometers are currently used onboard the \textit{Chandra X-ray Observatory} and \textit{XMM-Newton} as the main workhorses for X-ray spectroscopy with a resolution limit of 1000 ($\lambda /\delta\lambda$) and low effective area ($\lesssim100$ cm$^2$) over the same band.  Future goals of $>3000$ spectral resolving power and effective areas of $>1000$ cm$^2$ necessitate a new generation of high quality spectrometers capable of achieving these performance requirements \cite{McEntaffer13}. 

Off-plane reflection gratings are an attractive option for X-ray spectrometers.  They offer compact packing geometries, excellent grating efficiency, and the potential for very high resolving powers.  An array of off-plane gratings can be coupled with a set of nested Wolter-I optics (a primary parabolic mirror, followed by a secondary hyperbolic) to disperse a spectrum onto an imaging detector placed at the focal plane, typically a CCD camera \cite{McEntaffer10}.  The spectrum forms an arc of diffracted light in the shape of a cone, giving the common name for this type of diffraction---conical diffraction.

\begin{figure}
  \includegraphics[width=5.0in,height=2.63in]{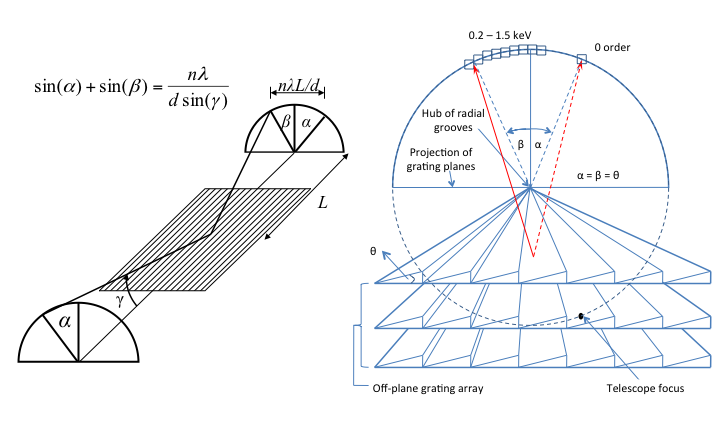}
\caption{\textit{Left} - The off-plane grating mount. \textit{Right} - Three gratings, placed many meters from the focus, are shown projected onto the focal plane to elucidate the nature of the arc of diffraction which is detected by an array of CCDs (depicted as squares).}
\label{OPGeom}
\end{figure}

To obtain future requirements of spectral resolving power and throughput, off-plane gratings require customized groove profiles.  Fig.\ \ref{OPGeom} depicts the grating geometry and outlines the necessary advancements.  The image on the left is the canonical off-plane geometry with light intersecting a ruled grating nearly parallel to the groove direction.  This creates an arc of diffraction at the focal plane with dispersion dictated by the displayed grating equation.  The image on the right is similar, but has the optical axis pointing out of the page.  The grating grooves are shown projected from the position of the gratings to a focal plane located several (typically $\sim8$) meters away.  High X-ray throughput requires high reflectivity and hence grazing incidence.  To increase the total collecting area, many gratings are stacked into an array.  Tight packing geometries are allowed because the cone angle of the diffracted light is roughly equal to the graze angle of the incoming light.

The effective area can be increased further by blazing the groove facets to a triangular profile that preferentially disperses light to one side of zero order.  This requires a smaller readout detector (or less detectors in an array) and thus increases the signal-to-noise in these orders.  The angle of the blaze on the grooves ($\theta$ in Fig.\ \ref{OPGeom}) is chosen to optimize diffraction efficiency toward the middle of the first order bandpass.  This, in turn, translates to optimized efficiencies at higher orders for shorter wavelengths.  The grating array is then rotated slightly about the grating normal resulting in an $\alpha$ for zero order at the focal plane that equals the $\beta$ of the optimized wavelength.  When $\alpha=\beta=\theta$ the array is in the Littrow configuration and is optimized for diffraction efficiency \cite{Cash83}. The similarity to the Littrow configuration in the in-plane diffraction sense can be seen by examining Fig.\ \ref{OPGeom} and setting $\alpha=\beta$.

The projection of the grooves in Fig.\ \ref{OPGeom} illustrates the radial distribution of grooves necessary to achieve high spectral resolving power \cite{Cash83}.  This convergence matches that of the telescope beam, thus maintaining a constant $\alpha$ over the grating.  This leads to a constant $\beta$ per wavelength at the focal plane and eliminates grating induced aberration due to the groove profile.  In other words, the converging rays from the Wolter I mirrors strike the grating at nearly the same angle with respect to the grooves at all points on the grating surface.

High groove density is another grating characteristic necessary for high spectral resolving power.  The baseline for future X-ray missions includes telescope optics with a $5-15$ arcsecond half power diameter (HPD).  The telescope beam can be sub-apertured \cite{Cash87} by the grating array to create a bowtie shaped line spread function (LSF) with a full width at half maximum (FWHM) of $\sim1-2$ arcsecond in the dispersion direction.  The dispersion of a grating measures the physical extent over which the spectrum is diffracted.  It is proportional to both the groove density and the distance between the grating array and the focal plane, the latter typically being called the throw.  Therefore, higher groove densities and/or longer throws increase dispersion, and therefore, spectral resolving power.  A throw of several meters ($\gtrsim5$) and a $\sim10$ arcsecond telescope HPD that can be sub-apertured down to $\sim1-2$ arcsecond translates to groove density requirements of $>5000$ grooves/mm.

Fig.\ \ref{OPGeom} also depicts the need for precision alignment within the off-plane grating array.  The grooves on each grating converge to a point at the center of the circle defined by the intersection of the cone of diffraction with the focal plane.  This focal circle is also coincident with the telescope focus and the zero order focus.  The gratings within an array are aligned such that all groove hubs are coincident.  Also, all grating surfaces must project to the diameter of the focal circle.  With these alignments achieved, the spectra from each grating overlap at the focal plane.

The off-plane mount provides a method for achieving the performance requirements of future soft X-ray spectroscopy missions.  However, developments in grating fabrication and alignment are necessary to ready this technology for flight.  First, off-plane gratings require custom profiles and higher groove densities in comparison to in-plane gratings.  Second, the alignment tolerances are tighter in comparison to transmission grating spectrometers.  Issues pertaining to the former are being addressed in a parallel study \cite{McEntaffer13}.  In this paper, we take the first step to address the latter by quantifying the off-plane alignment tolerances for a general spectrometer architecture.  First, we outline the mathematical formalism for analyzing the diffraction by an off-plane reflection grating.  Next, we define nominal alignment parameters using this formalism for a flight-like spectrometer.  Analytical alignment tolerances are obtained for all six degrees of freedom.  We verify these analytical calculations numerically using computer raytracing.  The alignment tolerances
are presented and examined for scalability with spectrometer focal length and spectral resolution requirement.  Finally, we investigate the dependence of diffraction efficiency on misalignments.

\section{Mathematical Formalism}
\label{sec:math}

\begin{figure}[ht]
\begin{minipage}[t]{0.5\linewidth}
\centering
\includegraphics[width=\textwidth]{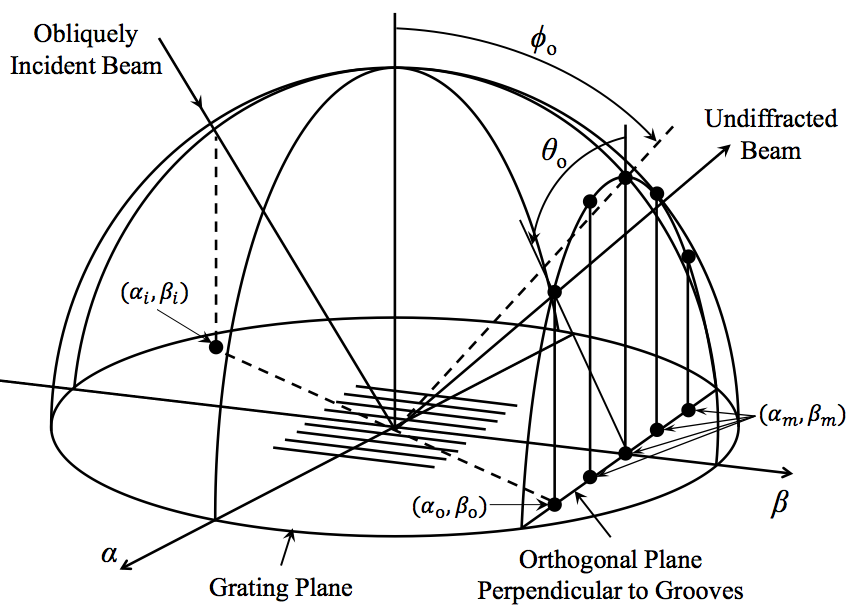}
\caption{The mathematical description of the off-plane grating geometry.  $\alpha_i,\beta_i$ describe the orientation of the incident beam, $\alpha_0,\beta_0$ describe the orientation of the specularly reflected beam, and $\alpha_m,\beta_m$ describe the orientation of the various orders of diffraction. Adapted from Harvey \& Vernold (1998).}
\label{fig:HarveyFigure}
\end{minipage}
\hspace{0.5cm}
\begin{minipage}[t]{0.5\linewidth}
\centering
\includegraphics[width=\textwidth]{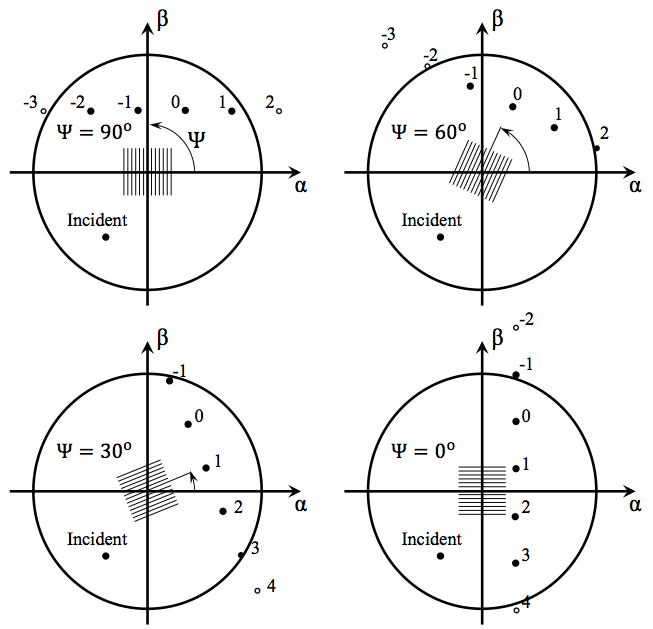}
\caption{Direction cosine diagrams for diffraction with various yaw angles $\Psi$. The location of zero order is set by the specular reflection of the incident beam. As $\Psi$ changes, the line of diffraction in cosine space rotates about zero order.  In this framework, the transition from classical in-plane diffraction to conical off-plane diffraction can be easily understood.  Adapted from Harvey \& Vernold (1998).}
\label{fig:HarveyFigure2}
\end{minipage}
\end{figure}

Harvey \& Vernold (1998) describe a convenient formalism for predicting the diffraction of light incident upon a parallel groove reflection grating for arbitrary grating orientation with respect to the incident beam.  This formalism makes use of direction cosines for the incident and diffracted rays, and the coordinate system used in this paper is shown in Fig.\ \ref{fig:HarveyFigure}.  $\alpha_i$ and $\beta_i$ are the direction cosines of the incident beam, $\alpha_0$ and $\beta_0$ are the direction cosines of the undiffracted, specularly reflected beam, and $\alpha_m$ and $\beta_m$ are the direction cosines of the diffracted beam of the $m$th order. In the figure, the grating grooves are aligned with the $\hat\beta$ axis.  The angle between the grating grooves and the $\hat\alpha$ axis is given by $\Psi$.  In Fig.\ \ref{fig:HarveyFigure}, $\Psi=90^\circ$ and the grooves are aligned with the $\hat\beta$ axis.  This formalism is completely general and reduces to that of Fig.\ \ref{OPGeom} in the limit of $\Psi \rightarrow 90^\circ$. Note that these coordinates are not related to the angles $\alpha$ and $\beta$ in Fig.\ \ref{OPGeom}.

The angular coordinates in real space are $\theta$ and $\phi$, where $\phi$ is the polar angle with respect to the grating normal and $\theta$ is the subsequent rotation angle about the $\hat\beta$ axis, with the same subscripts associated with $\alpha$ and $\beta$.  By solving for the direction cosines of the $m$th diffracted beam, one can solve for the angle $\theta_m$ ($\phi_m = \phi_0$) and thus find the direction vector describing the diffracted beam in real space.  We define our real space coordinate system as $\hat x=\hat\alpha$, $\hat y =\hat\alpha\times\hat\beta$, and $\hat z = \hat\beta$. This formalism says nothing about the efficiency of the diffracted orders, which will be addressed in \S \ref{sec:Area}.

Using these coordinates, the equations giving the diffracted directions for arbitrary beam incidence and grating orientation are \begin{equation} \label{eq:cosines}\begin{array}{c} \alpha_m + \alpha_i = (m\lambda/d){\rm sin}\Psi \\ \beta_m + \beta_i = -(m\lambda/d){\rm cos}\Psi \end{array}, \end{equation} where $\alpha_m = {\rm sin}\theta_m{\rm cos}\phi_m$, $\beta_m={\rm sin}\phi_m$, $\alpha_i = -{\rm sin}\theta_0{\rm cos}\phi_0$, and $\beta_i = -{\rm sin}\phi_0$.  Fig.\ \ref{fig:HarveyFigure2} shows the various diffracted orders in direction cosine space for various grating orientations.  The zero order, specularly reflected beam location is fixed based on the incident beam's location, the spacing of the various orders is dictated by the wavelength $\lambda$ and groove period $d$, and the orientation of the line of diffraction in direction cosine space is dictated by the grating orientation $\Psi$.  Orders lying outside the $\alpha^2+\beta^2=1$ circle in direction cosine space are so-called evanescent orders, and are not observed in real space.

Note that the correct oblique beam alignment is achieved in practice by setting the incidence angle $i$ and an effective yaw angle $\Psi_{\rm eff}$ (groove direction) of a grating with respect to the nominal optical axis of the mirror pair to which it is aligned.  In other words, the grating position is adjusted to a stationary optical axis, rather than vice versa.  This can be understood in a reference frame where $\theta_0^\prime=0$ and $\phi_0^\prime=i$.  The beam impact geometry defined by the original $\theta_0$, $\phi_0$, and $\Psi$ defined above is preserved using $i={\rm sin}^{-1}({\rm cos}\phi_0{\rm cos}\theta_0)$ and $\Psi_{\rm eff}=\Psi+{\rm tan}^{-1}({\rm sin}\theta_0{\rm tan}\phi_0)$.

\section{Assumptions}
\label{sec:Assumptions}

\begin{figure}
\centering
  \includegraphics[height=8cm]{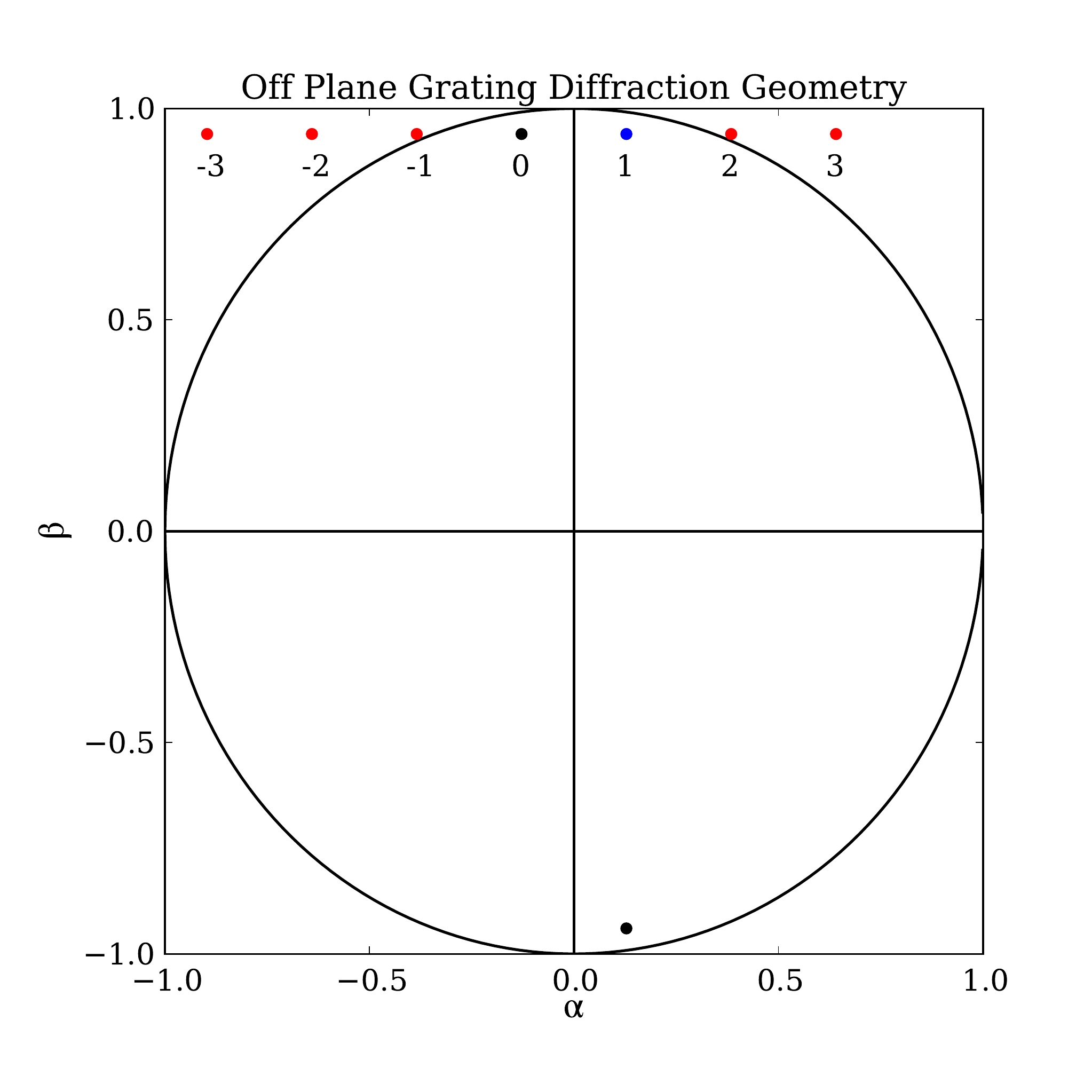}
\caption{A representative diffraction cosine diagram for the assumptions in \S \ref{sec:Assumptions}. The beam incidence angles and groove period have been adjusted in order to spread the diffraction orders out for readability.  Evanescent diffraction orders are shown in red, while the non-evanescent first diffraction order is shown in blue.  For our assumptions the diffraction orders would be clustered together near the top of the diagram such that they would not be visually distinguishable.}
\label{fig:DiffGeom}
\end{figure}

For our initial grating alignment calculations, we will be using radial gratings with a groove period of $d=160$ nm at an 8 m distance from the hub.  Longer wavelength light is diffracted at larger angles, requiring tighter alignment tolerances. Thus, we assume a wavelength of 4.1 nm, corresponding to an energy of 0.3 keV (the low end of our desired energy range).  We take a characteristic initial beam alignment with an incidence angle $i=88.5^\circ$ and $\theta_0=-18^\circ$, where ${\rm sin}i={\rm cos}\phi_0{\rm cos}\theta_0$.  This $\theta_0$ will optimize the grating for a first order Littrow configuration at a wavelength of roughly 4 nm.   The sign of $\theta_0$ is arbitrary; one of the first order beams is diffracted into evanescence, while the other is available for spectroscopy.  We set our nominal yaw to $\Psi=90^\circ$. Finally, we assume a flat focal plane positioned a distance $L=8$ m from the nominal beam impact point (the origin in Fig.\ \ref{fig:HarveyFigure}) along the $\hat z$ axis and parallel to the $xy$ plane.  This distance is typical of X-ray grating spectrometer architectures recently studied by NASA \cite{Bautz,McEntaffer11}.  A direction cosine diagram illustrating our assumptions is shown in Fig.\ \ref{fig:DiffGeom}.  For our assumptions, the diffraction orders would be clustered near $\alpha=0,\beta=1$ and would not be visually distinguishable. We therefore set $\phi_0$, $\theta_0$, and $d$ to 70$^\circ$, $-22^\circ$, and 16 nm, respectively, in order to make the diagram readable.  The diagram is qualitatively consistent with our assumptions.  Note that the only diffraction order not in evanescence is the first and the location of the first diffraction order is in the Littrow configuration ($\alpha_1=\alpha_i$ and $\beta_1=-\beta_i$).

\section{Spectral Resolution Spot Shift Requirements}
\label{sec:spr}

Reflection grating spectroscopy turns the spectral resolution problem into a spatial resolution problem.  Photons will be dispersed in the $\hat x$ direction based on their wavelength.  Thus, we map a photon's $x$ position on the detector to wavelength or energy regardless of its $y$ position as evident in the upper left panel of Fig.\ \ref{fig:HarveyFigure2}.  For a single grating, the spectral resolution is then dictated by the point spread function of the diffracted image in the $\hat x$ direction.  With current sub-apertured Wolter I optics produced by Zhang et al.\ (2012), the line spread function (LSF) in the dispersion direction ($\hat x$) is approximated by a line with a 1 arcsecond full width at half maximum (FWHM).  The line width is oriented in the dispersion direction so that energy resolution is dictated by the 1 arcsecond spread.  For our throw length ($L$), this is equivalent to roughly 40 $\mu$m.  Thus, an estimate of our spectral bin size is 40 $\mu$m.

When a second grating is integrated into the spectrometer array, it must be aligned such that its arc of diffraction coincides with that of the first grating.  Misalignments, both angular and translational, will cause the diffracted beam of a given order and wavelength to shift.  Our initial goal is to limit this shift in the $\hat x$ direction to less than 40 $\mu$m (the spectral bin size) at the detector plane.  As a first step, we will calculate the maximum allowable misalignment for each independent degree of freedom.  The 40 $\mu$m spot shift limit is somewhat arbitrary; the actual spot shift requirement for an instrument would flow down from a top level spectral resolution requirement. This will be addressed in a future paper investigating both coupled alignment tolerances and optimization of spectral resolution and diffraction efficiency.

To calculate the effect a given misalignment has on a photon's $x$ position at the focal plane, we must express $x$ as a function of $\alpha_m$ and $\beta_m$.  Let $\Theta$ be the angle between the $\hat z$ (or $\hat\beta$) axis and the projection of the diffracted beam onto the $xz$ plane.  The distance between the beam impact point and the focal plane is $L$, thus the beam's $x$ position at the focal plane is $L{\rm tan}\Theta$.  $\Theta$ can also be expressed as arctan$(\alpha_m/\beta_m)$, leading to $x=L\alpha_m/\beta_m$.  Finally, to calculate the shift in $x$ position of the $m$th diffracted beam due to a misalignment, initial (subscript 0) and final (subscript 1) direction cosines can be used to obtain \begin{equation}\label{eq:spot} \Delta x=L_1(\alpha_{m,1}/\beta_{m,1}) - L_0(\alpha_{m,0}/\beta_{m,0}),\end{equation} where $L_0 = 8$ m and $L_1$ is the final throw length after a possible shift in the beam impact location due to translations of the grating.

\section{Analytical Alignment Tolerances}
\label{sec:SpecLimits}

In this section, Equations \ref{eq:cosines} and \ref{eq:spot} are used to calculate the maximum misalignment such that $\Delta x\leq 40~\mu$m.  For each degree of freedom, perfect alignment is assumed for the other five degrees of freedom.

\subsection{Yaw}

\begin{figure}[ht]
\begin{minipage}[t]{0.5\linewidth}
\centering
\includegraphics[width=\textwidth]{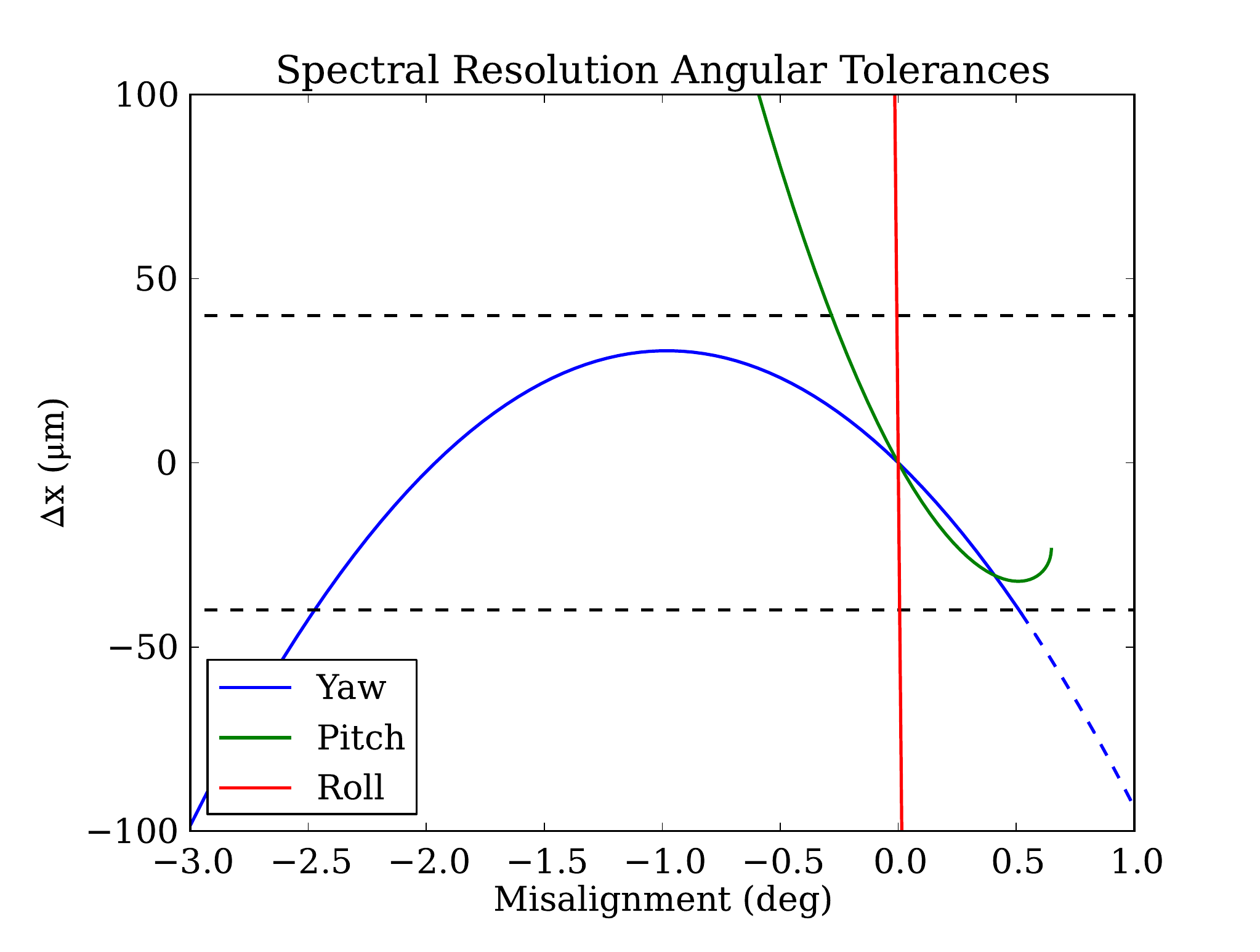}
\caption{Shifts in the $x$ position of the diffracted beam at the focal plane as a function of angular misalignment.  Assumptions are as in \S \ref{sec:Assumptions}.  A dashed line along a curve indicates a transition into evanescence. The horizontal black dashed lines indicate the spectral resolution constraints on $\Delta x$.}
\label{fig:AngleEffect}
\end{minipage}
\hspace{0.5cm}
\begin{minipage}[t]{0.5\linewidth}
\centering
\includegraphics[width=\textwidth]{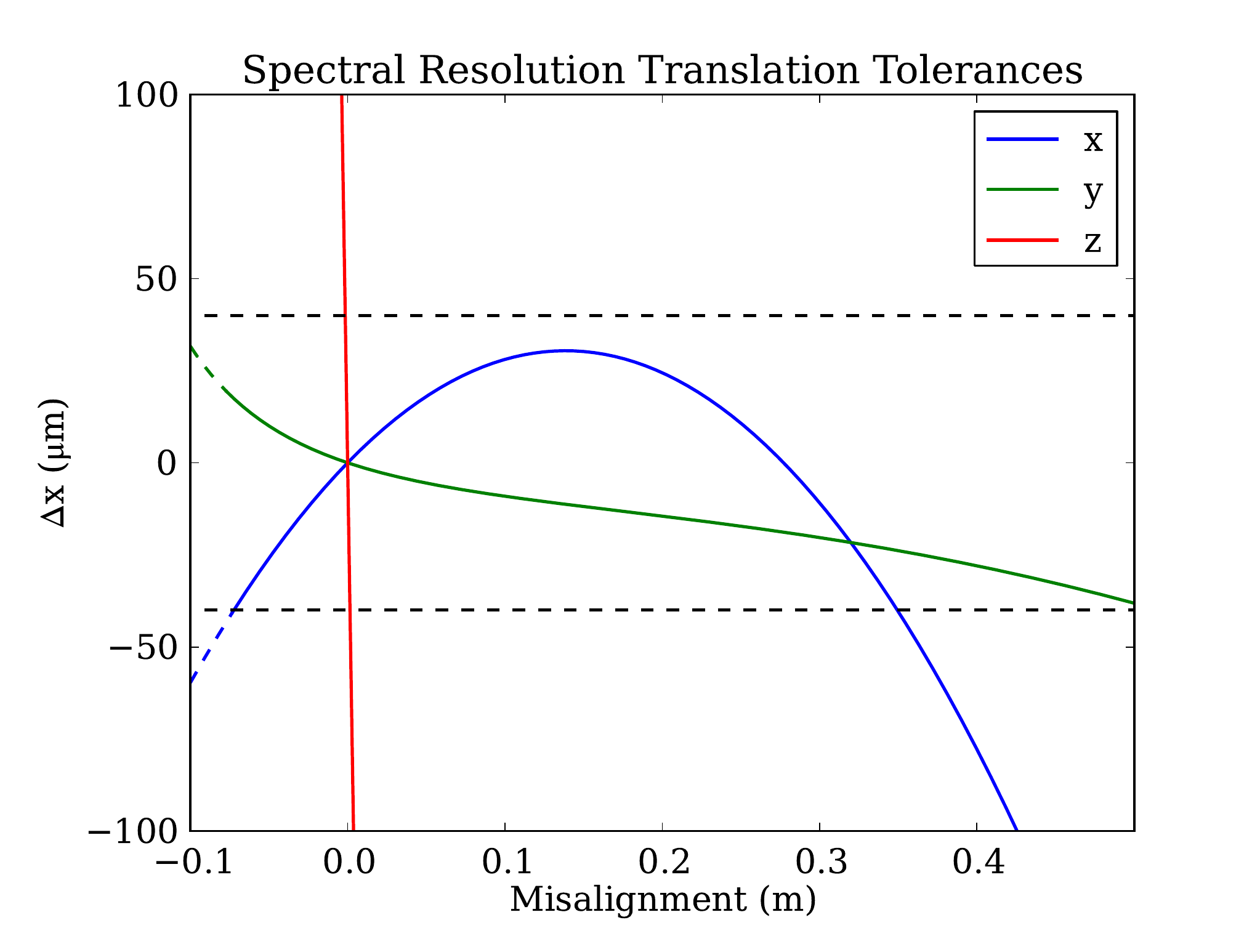}
\caption{Shifts in the $x$ position of the diffracted beam at the focal plane as a function of translational misalignment.  Assumptions are as in \S \ref{sec:Assumptions}.  A dashed line along a curve indicates a transition into evanescence. The horizontal black dashed lines indicate the spectral resolution constraints on $\Delta x$.}
\label{fig:TransEffect}
\end{minipage}
\end{figure}

As the yaw angle of the grating ($\Psi$) is changed, the line of diffraction in direction cosine space is rotated about zero order as in Fig.\ \ref{fig:HarveyFigure2}.  This inherently asymmetric effect is shown in Fig.\ \ref{fig:AngleEffect}.  This figure shows the positional dependence ($\Delta x$) of a first order, 4.1 nm spectral line on alignment errors in the rotational degrees of freedom.  As $\Psi$ is rotated in the positive (counterclockwise) direction in Fig.\ \ref{fig:HarveyFigure2}, first order is rotated up toward evanescence (out of the $\alpha^2 + \beta^2$ unit circle), which it eventually reaches at roughly $\Psi=0.5^\circ$.  As $\Psi$ is rotated in the negative (clockwise) direction, at first the $x$ position increases because $\beta_{m,1}$ is increasing while $\alpha_{m,1}$ is nearly constant with small yaw rotations.  Eventually, $\alpha_{m,1}$ begins to rapidly decrease as $\Psi$ continues to decrease, resulting in an inflection point in the $x$ position with respect to yaw angle.  The bounds on yaw are constrained by our spectral requirement (horizontal dashed lines in Fig.\ \ref{fig:AngleEffect}) at $-2.47^\circ$ and $+0.52^\circ$.  The first order beam becomes evanescent shortly after the upper bound is surpassed.

\subsection{Pitch}

Changing the pitch angle results in a change in both $\theta_0$ and $\phi_0$, because the incident beam is rotated about the $\hat x$ ($\hat \alpha$) axis.  The new beam incidence is found by applying the appropriate rotation matrix $R_x$, which results in a first order diffraction direction with respect to the new grating plane.  Then, the inverse rotation $R^{-1}_x$ must be used to convert the diffraction direction back in the original coordinate system of the grating with zero pitch offset.  The initial and final direction cosines are then used to calculate $\Delta x$. These results are shown in Fig.\ \ref{fig:AngleEffect}, where a positive pitch angle indicates a more glancing incidence angle.  As pitch is increased, the entire line of diffraction as in Fig.\ \ref{fig:HarveyFigure2} moves up toward evanescence, which first order (4.1 nm) reaches at $0.65^\circ$.  A decrease in pitch causes the line of diffraction to move away from evanescence, and creates more spectral displacement in our focal plane.  Spectral resolution sets the lower bound on pitch at $-0.28^\circ$.

\subsection{Roll}

Changing the roll angle ($\theta_0$) produces a rotation of the diffracted orders at the detector plane about the $\hat z$ ($\hat \beta$) axis, at least for small roll angles at a glancing beam incidence.  The shift of the diffraction spots in the $\hat x$ direction due to this rotation is given by $L{\rm tan}i{\rm sin}R$, where $i$ is the incidence angle and $R$ is the roll angle.  Equating this to 40 $\mu$m yields a roll angle tolerance of $\pm 40$ arcsec.  An additional effect, which is important for our calculations, is the change in diffraction due to the change in $\theta_0$.  Similar to the preceding section, we analyze roll effects by using a rotation matrix $R_z$ to determine the beam incidence with respect to the rolled grating.  The first order diffracted direction is then rotated back into the original coordinate system using $R^{-1}_z$.  The results are shown in Fig.\ \ref{fig:AngleEffect}, and is approximately linear in the relevant angular range.  Roll is constrained by spectral resolution at $\pm 21.6$ arcsec.  Note that this is tighter than the tolerance predicted simply by rotating the diffraction spots about $\hat z$.

\subsection{$\hat x$ Translations}

Translating the grating in the $\hat x$ ($\hat \alpha$) direction results in a change in yaw angle.  This occurs due to the radial grooves which are all directed toward the same point in the focal plane (i.e.\ they follow the cone angle of the incident beam).  For large $\phi_0$, there is a negligible change in groove spacing $d$.  A shift $\delta x$ results in an effective yaw equal to arctan$(\delta x/L)$.  One can then use the results from the yaw analysis to determine translational bounds in the $\hat x$ direction of 73.3 mm and 349 mm.  As with the rotational degrees of freedom, changes in the first order diffraction spot in the focal plane are also directly calculated as a function of $\hat x$ grating translations using Equations \ref{eq:cosines} and \ref{eq:spot} and are shown in Fig.\ \ref{fig:TransEffect}.

\subsection{$\hat z$ Translations}

For our beam geometry, $\hat z$ translation of the grating changes the grating period $d$ with a negligible change in yaw angle.  The groove period is linear with distance along the $\hat z$ direction, and can be expressed as a function of the nominal period and the nominal throw length: $d(z)=z(d_0/L_0)$.  Taking our nominal $z$ position to be 8 m, we can write the groove period as a function of $\hat z$ grating translation $\delta z$ as $d(\delta z)=(8~{\rm m}+\delta z)(160~{\rm nm}/8~{\rm m})$.  Using this expression in Eq.\ \ref{eq:cosines} to compute direction cosines for use in Eq.\ \ref{eq:spot} leads to $\hat z$ translation bounds of $\pm 1.51$ mm.  The results of this calculation are shown in Fig.\ \ref{fig:TransEffect} and demonstrate that the effect is approximately linear in the range of interest.

\subsection{$\hat y$ Translations}

$\hat y$ translations will move the point of incidence, changing both the throw length and the groove period in the process.  The total shift in the beam impact location on the grating is $sl=\delta y/{\rm tan}i=\delta y/{\rm tan}({\rm sin}^{-1}({\rm cos}\theta_0{\rm cos}\phi_0))\simeq\delta y/({\rm cos}\theta_0{\rm cos}\phi_0)$ (for large $\phi_0$).  The $\hat z$ component of this shift can be written $sz = sl{\rm sin}\phi_0=\delta y {\rm tan}\phi_0/{\rm cos}\theta_0$.  Then, the throw length goes to $L+sz$ and the groove period goes to $(L+sz)(d_0/L)$ as in the previous section.  This leads to a change in the location of the diffracted spot of $\Delta x=(L+sz)({\rm sin}\theta_0{\rm cos}\phi_0 +m\lambda L/(d_0(L+sz))) - L({\rm sin}\theta_0{\rm cos}\phi_0 + m\lambda/d_0)=\delta y{\rm tan}\theta_0{\rm sin}\phi_0$.  However, there is also a $\hat x$ shift in the beam impact location $sx=sl{\rm sin}\theta_0{\rm cos}\phi_0=\delta y {\rm tan}\theta_0$.  $sx$ is in the opposite direction of $\Delta x$, and in the limit of large $\phi_0$ the two effects cancel.  These effects are calculated using our alignment assumptions and shown in Fig.\ \ref{fig:TransEffect}, where the upper limit of 0.5 m is due to a spectral resolution cutoff and the lower limit of 76.9 mm is due to the first order becoming evanescent.

\section{Effective Area Considerations}
\label{sec:Area}

\begin{figure}[ht]
\begin{minipage}[t]{0.5\linewidth}
\centering
\includegraphics[width=\textwidth]{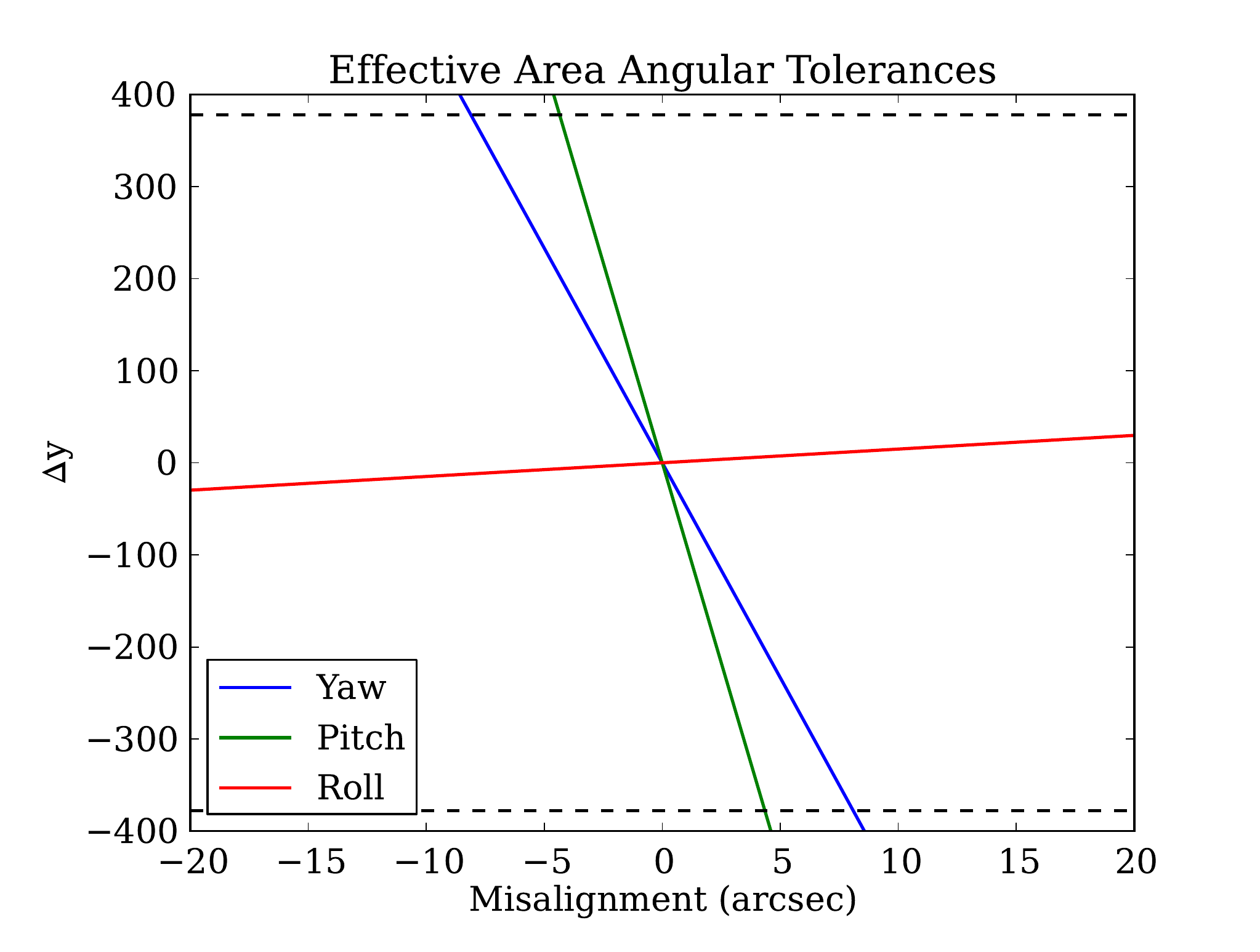}
\caption{Shifts in the $y$ position of the diffracted beam at the focal plane as a function of angular misalignment.  Assumptions are as in \S \ref{sec:Assumptions}. The horizontal black dashed lines indicate the effective area constraints on $\Delta y$.}
\label{fig:AreaAngles}
\end{minipage}
\hspace{0.5cm}
\begin{minipage}[t]{0.5\linewidth}
\centering
\includegraphics[width=\textwidth]{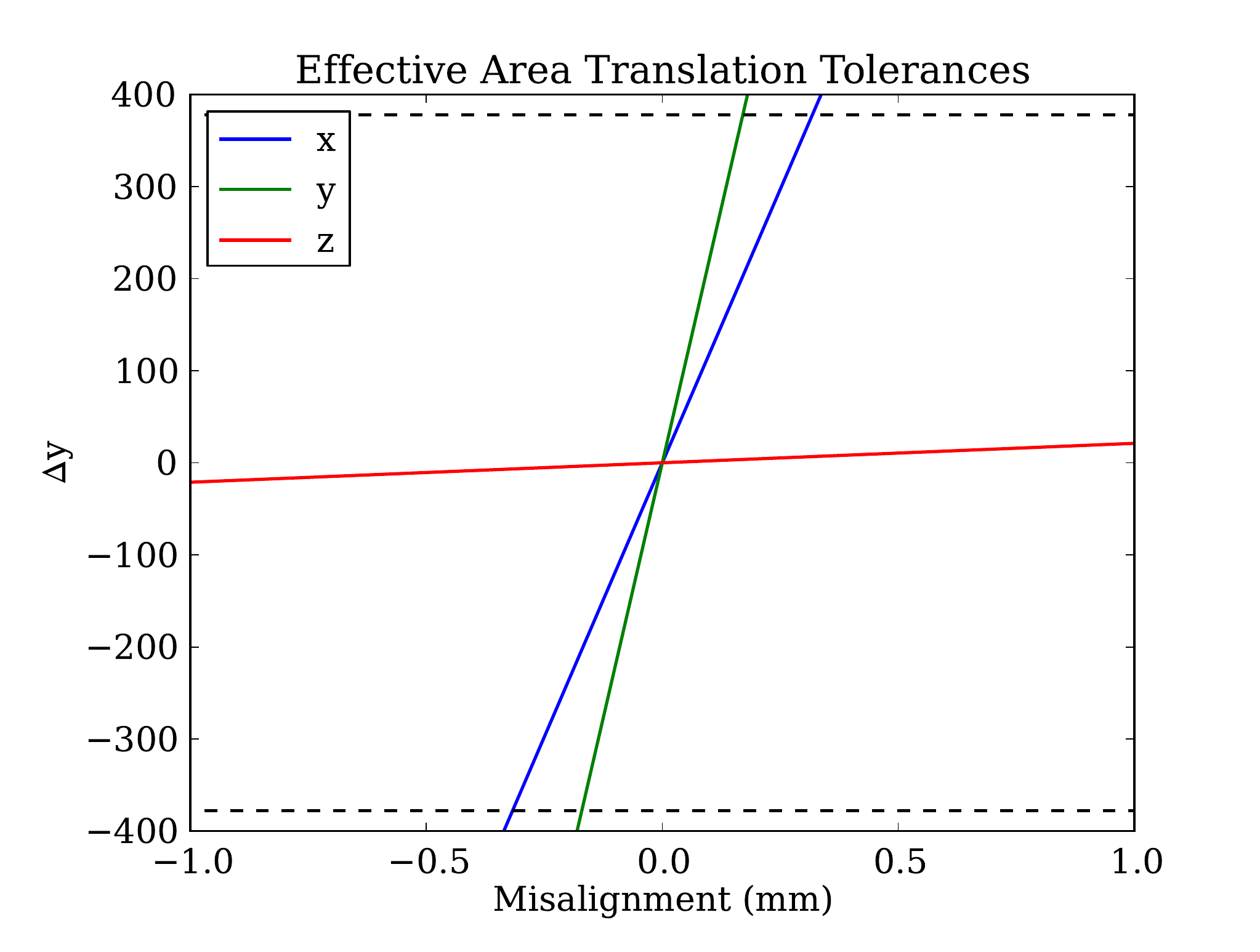}
\caption{Shifts in the $y$ position of the diffracted beam at the focal plane as a function of translational misalignment.  Assumptions are as in \S \ref{sec:Assumptions}. The horizontal black dashed lines indicate the effective area constraints on $\Delta y$.}
\label{fig:AreaTranslations}
\end{minipage}
\end{figure}

A broadening of the arc of diffraction in the $\hat y$ direction will result in an effective area loss due to the deposited charge being spread over a greater number of CCD pixels.  This increased noise can be limited by reducing the alignment tolerances such that the diffraction spot always rests within the 10 arcsecond half power diameter (HPD) of the telescope point spread function (PSF) in the $\hat y$ direction.  With our 8 m throw length, this translates into a 378 $\mu$m $y$ shift in the CCD plane.  As in \S \ref{sec:spr}, this limit is somewhat arbitrary.  This analysis is conceptually similar to the above spectral resolution analysis, with the shift \begin{equation}\label{eq:dy}\Delta y=L_1{\rm tan}({\rm arcsin}(\gamma_{m,1}))-L_0{\rm tan}({\rm arcsin}(\gamma_{m,0})),\end{equation} where $\gamma$ is the direction cosine in the $\hat \alpha \times \hat \beta$ direction.  The third direction cosine can easily be calculated using $\gamma=\sqrt{1-\alpha^2-\beta^2}$.  Analyses analogous to those of \S \ref{sec:SpecLimits} were carried out and are presented in Figs. \ref{fig:AreaAngles} and \ref{fig:AreaTranslations}.  The tolerances for yaw, pitch, $\hat x$, and $\hat y$ translation are tightened to $\pm 7.9$ arcsec, $\pm 4.3$ arcsec, $\pm 317~\mu$m, and $\pm 170~\mu$m, respectively.  The new tolerances produced for roll and $\hat z$ translations are looser than those calculated in \S \ref{sec:SpecLimits}. The results are summarized in Table 1 (see \S \ref{sec:Requirements}).

\section{Raytracing Verification}

\begin{figure}
  \includegraphics[width=\textwidth]{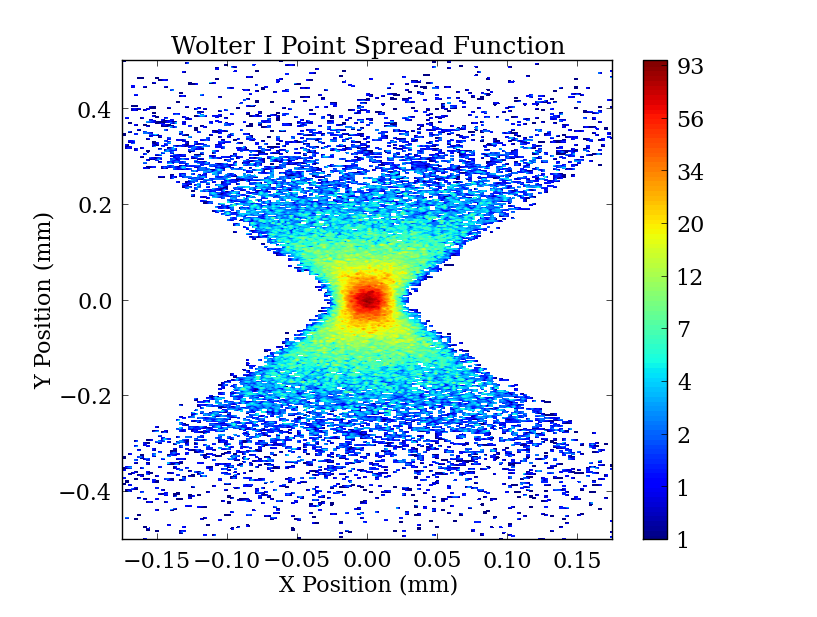}
\caption{The PSF obtained from our sub-apertured Wolter I raytracing code.  It is approximated by a line with a 1 arcsecond (40 $\mu$m) Gaussian spread in the dispersion ($\hat x$) direction.  The spread in the orthogonal direction is dominated by microroughness induced scatter.  This is characteristic of the PSF of modern mirrors produced by Zhang et al.\ (2012).}
\label{fig:PSF}
\end{figure}

The above tolerances were verified by raytracing using Interactive Ray Trace \copyright~(IRT; Parsec Technology, Inc.).  A standard Wolter I telescope geometry was used with mirror parameters as defined in van Speybroeck \& Chase (1972).  Our mirror parameters were $L_h=L_p=200$ mm, $r_0=244.5$ mm, and $Z_0=8400$ mm.  We included a 25 mm gap between the primary and secondary mirrors.  We then placed a $100\times100$ mm$^2$ radial grating with a 50 mm gap in the axial direction between the bottom edge of the secondary mirror and the top edge of the grating.  The $1.5^\circ$ grating incidence angle was defined by the angle between the grating surface and the unit vector from the center of the secondary mirror to the focus. The grating center was placed to intersect this unit vector.  The grating hub was located a distance 8088 mm from the grating center.  Scatter based on the theory of Beckmann \& Spizzichino (1987) was added to the photons at the primary mirror to account for microroughness.  To produce the LSF in the dispersion direction obtained by Zhang et al.\ (2012), a 1 arcsecond Gaussian spread was added along the $\hat x$ axis.  As in \S \ref{sec:SpecLimits}, we assume a 1 arcsecond FWHM in the dispersion direction and a 10 arcsecond HPD in the orthogonal direction.  The photons were traced to the plane orthogonal to the grating and intersecting the grating hub.  The resultant PSF is shown in Fig.\ \ref{fig:PSF}.

\begin{figure}[ht]
\begin{minipage}[t]{0.5\linewidth}
\centering
\includegraphics[width=\textwidth]{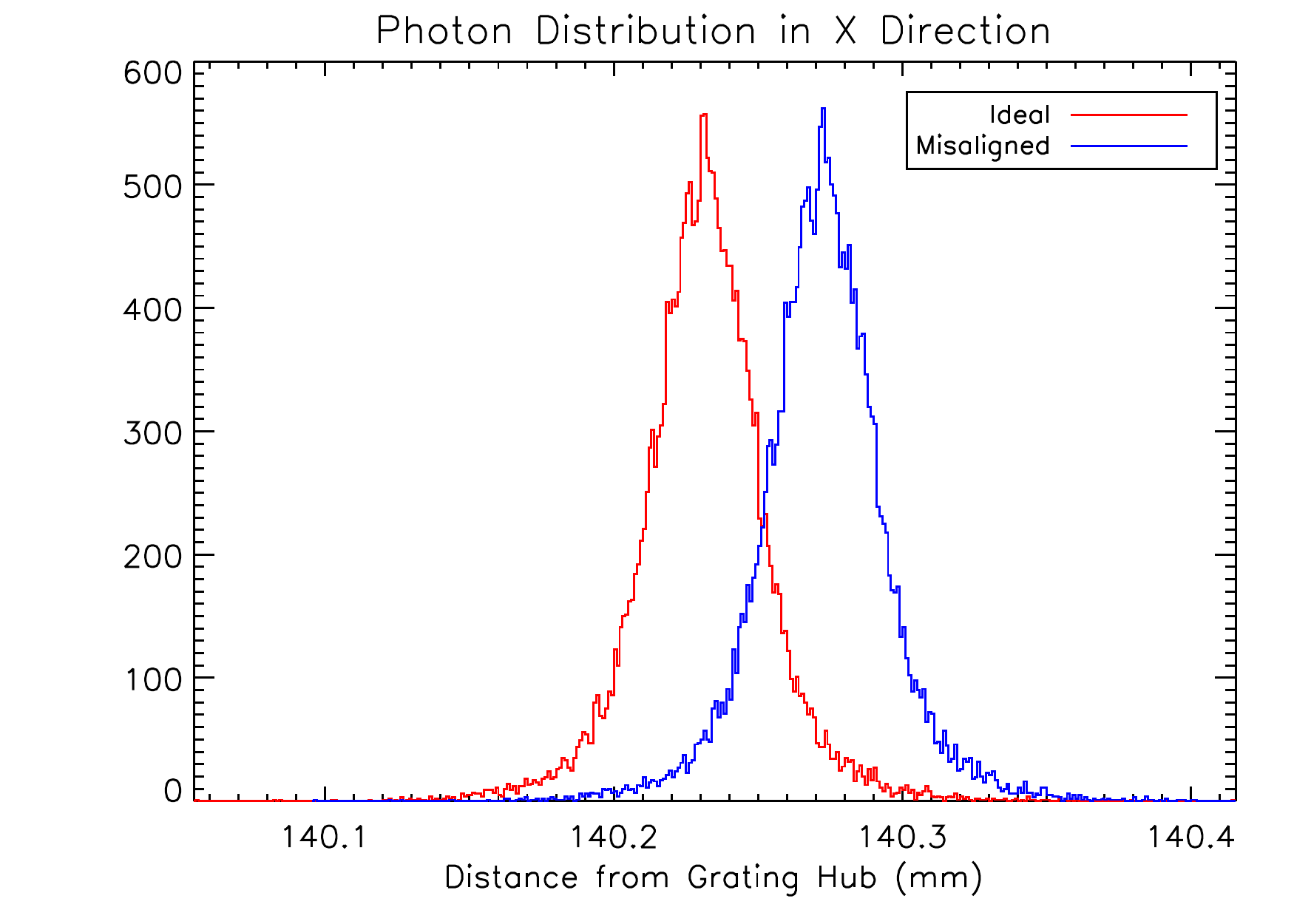}
\caption{The distribution of counts from the PSF of Fig.\ \ref{fig:PSF} in the $\hat x$ direction. The ideal distribution is obtained for a nominally aligned grating.  The misalignment distribution is obtained after the roll tolerance of 22 arcseconds is reached.  The 40 $\mu$m (1 arcsecond) Gaussian spread is caused by mirror figure error.}
\label{fig:XHist}
\end{minipage}
\hspace{0.5cm}
\begin{minipage}[t]{0.5\linewidth}
\centering
\includegraphics[width=\textwidth]{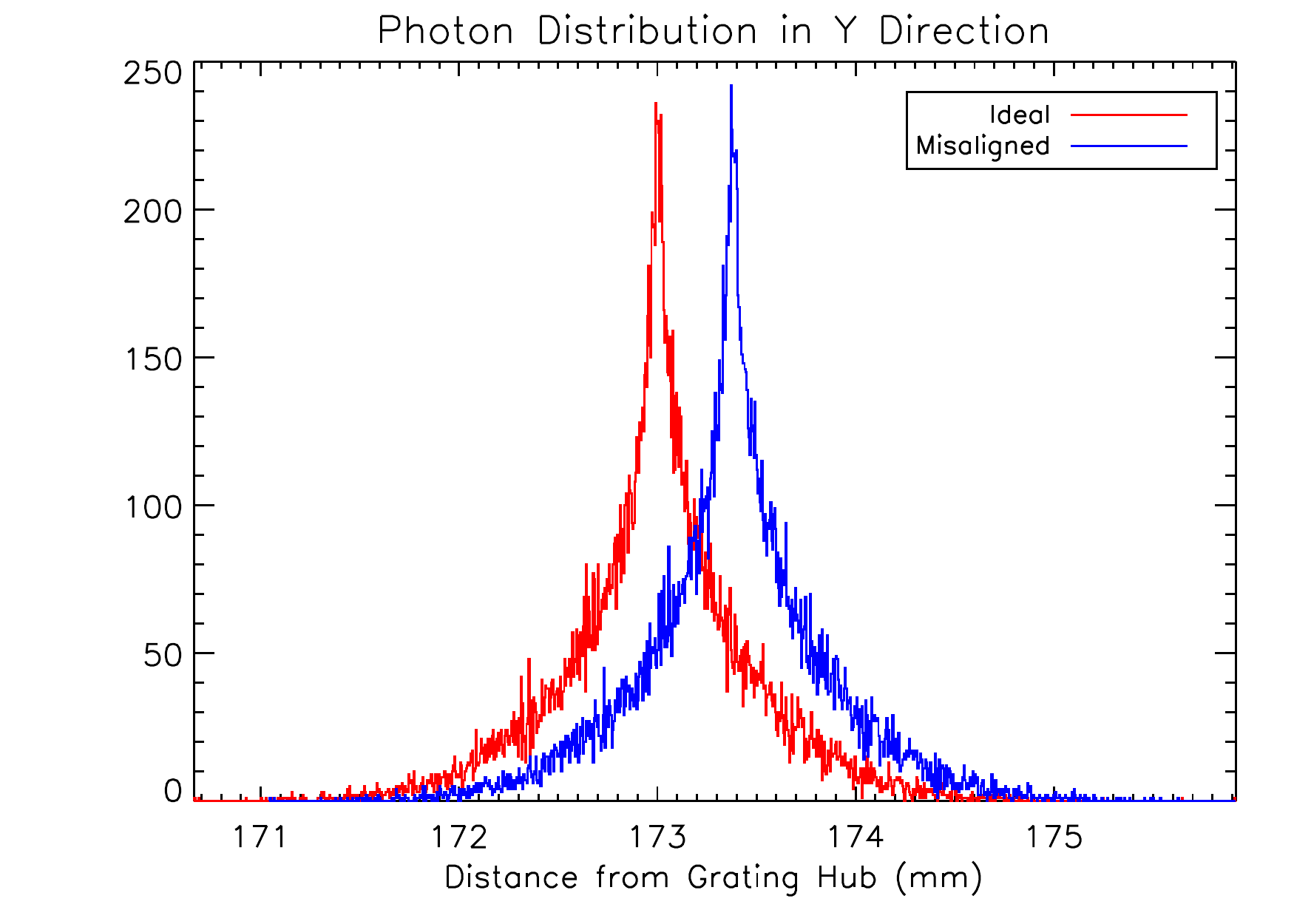}
\caption{The distribution of counts from the PSF of Fig.\ \ref{fig:PSF} in the $\hat y$ direction. The ideal distribution is obtained for a nominally aligned grating.  The misalignment distribution is obtained after the $\hat y$ translation tolerance of 170 $\mu$m is reached.  The 378 $\mu$m ($\sim 10$ arcseconds) spread is caused by microroughness induced scatter.}
\label{fig:YHist}
\end{minipage}
\end{figure}

To verify the analytical tolerances, a misalignment was introduced and photons were traced to the CCD plane and compared to the nominal PSF (Fig.\ \ref{fig:PSF}).  The misalignment was increased until one of three tolerance conditions were violated: 1) The mean $x$ position was $>40~\mu$m from that of the nominal PSF, 2) The mean $y$ position was $>378~\mu$m from that of the nominal PSF, and 3) More than 10\% of the rays were cut off due to evanescence.  Histograms of the $x$ positions for a 22 arcsecond roll misalignment and $y$ positions for a 170 $\mu$m $\hat y$ translation are shown in Figs.\ \ref{fig:XHist} and \ref{fig:YHist}.  All six degrees of freedom produced faults via either condition 1 or 2, consistent with \S \ref{sec:SpecLimits}.

\section{Alignment Requirements and Scalability}
\label{sec:Requirements}

Table 1 summarizes our verified alignment tolerances.  The results plotted in \S \ref{sec:SpecLimits} indicate linearity of the alignment tolerances with respect to spot shift requirements.  To investigate the limits of this linearity, lines were fit to the limiting tolerance curves within the $|\Delta x|<40~\mu$m or $|\Delta y|<378~\mu$m bounds.  The percent difference between the fitted lines and the full misalignment curves were then limited to $<1\%$.  The $\Delta x$ or $\Delta y$ values at which this condition was violated were taken as the linearity limits.  This occurred at more than 9 times the $40~\mu$m or $378~\mu$m bounds for all degrees of freedom.  The factor by which $\Delta x$ or $\Delta y$ can be increased before the linearity is broken is listed in Table 1 as the linearity factor.  In other words, increasing the $\Delta x$ and $\Delta y$ limits by a factor of 9 to 360 $\mu$m and 3.4 mm, respectively, results in all of the tolerances being increased by that same factor. Thus, these tolerances are linear for less stringent spectral resolution and effective area requirements up to a factor of about 9, assuming a fixed focal length.  Tightening the spectral resolution and effective area requirements clearly maintains linearity on the tolerances up to any factor.  These tolerances are also predicted to scale linearly with the focal length based on Eq.\ \ref{eq:cosines} and Eq.\ \ref{eq:spot}.

\begin{table}[h]
\centering
\label{tab:Limits}
\begin{tabular}{c | c | c | c}
\multicolumn{4}{c}{Table 1: Analytical Alignment Tolerance Summary} \\
\hline \hline
Misalignment & Tolerance & Limiting Effect & Linearity Factor \\
\hline
Yaw & $\pm7.9$ arcsec & Effective Area & 9 \\
Pitch & $\pm4.3$ arcsec & Effective Area & 106 \\
Roll & $\pm21.6$ arcsec & Spectral Resolution & 174 \\
$\hat x$ & $\pm317~\mu$m & Effective Area & 9\\
$\hat y$ & $\pm170~\mu$m & Effective Area & 42 \\
$\hat z$ & $\pm1.51$ mm & Spectral Resolution & 52 \\
\hline\hline
\end{tabular}
\end{table}

\section{Diffraction Efficiency}

As mentioned in \S \ref{sec:math}, the formalism used in this paper does not account for changes in diffraction efficiency.  If the diffraction efficiency per grating were to change appreciably due to misalignments, calculating the effective area of the grating spectrometer would have to take this into account.  In fact, this would also greatly complicate the energy response function of the spectrometer: the precise alignment of each mirror pair and grating would need to be known to compute the on-axis energy response function, and an off-axis source could potentially change the response function in a significant manner.  Fortunately, we have performed efficiency simulations that show this is not a concern for our expected misalignments.

We use the commercial software PCGrate-S(X) v.6.1 \copyright~(I.I.G.\ Inc.) to compute our efficiencies, as the dependence of diffraction efficiency on beam geometry and wavelength is a complicated computational problem (see Neviere \& Popov 1998 for a review).  The software works by solving a system of integral equations over the periodic groove boundary.  We assume a grating with an $18^\circ$ blaze angle and a 160 nm period.  The groove profile is right triangular, and the groove material is gold.  The nominal $\theta_0$ and $\phi_0$ are as in \S \ref{sec:Assumptions}.  The normal computation mode is used with the standard options, and we obtain normal accuracy conditions (e.g.\ relative efficiency summed over all orders is 1) for all calculations reported in this paper.

\begin{figure}
  \includegraphics[width=\textwidth]{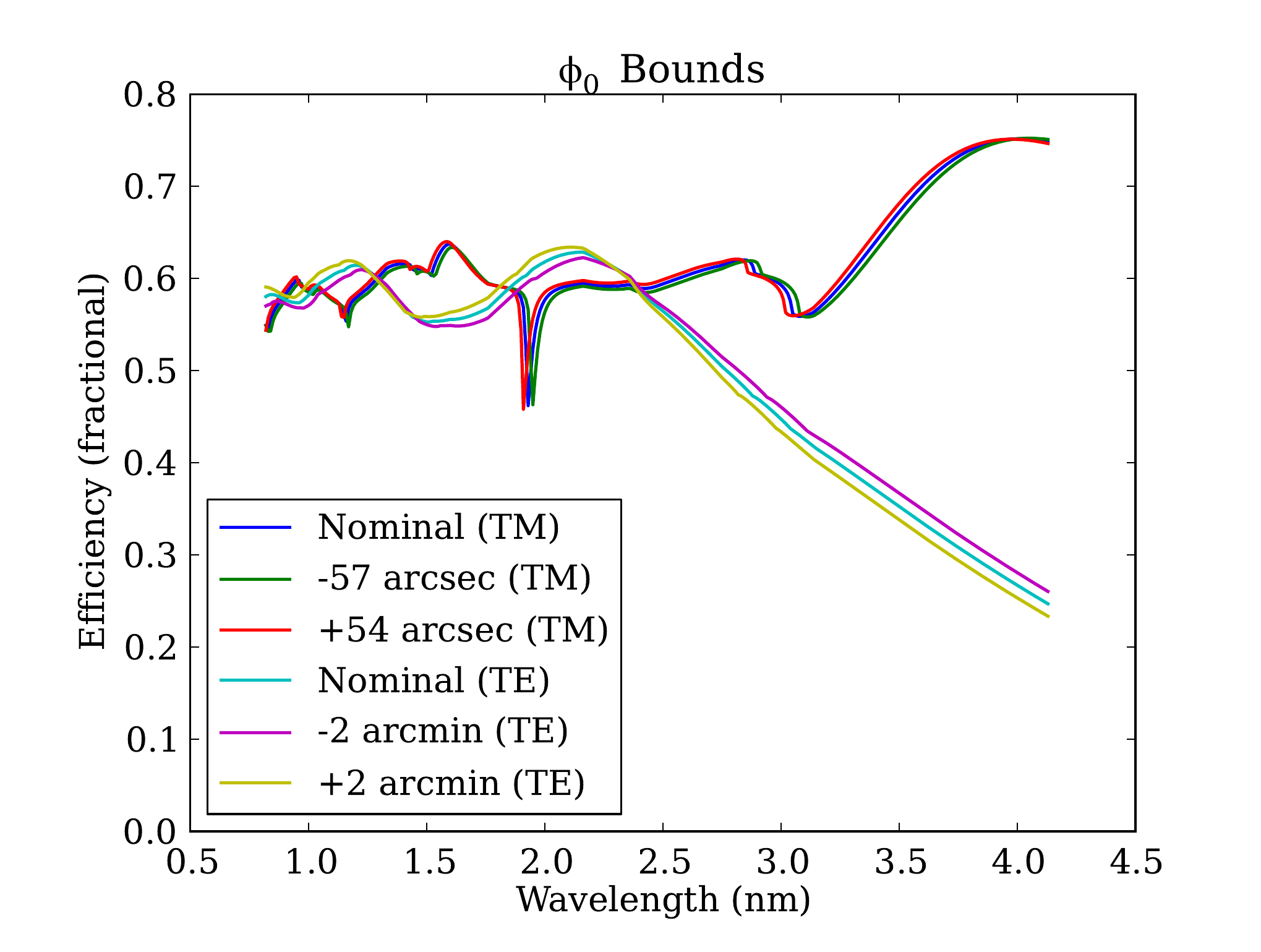}
\caption{Diffraction efficiencies for both transverse electric (TE) and transverse magnetic (TM) polarizations for a nominal ($88.4228^\circ$) and misaligned $\phi_0$. For our angular misalignment tolerances, the maximum change in $\phi_0$ is $\pm4$ arcseconds (pitch).}
\label{fig:PhiBounds}
\end{figure}

\begin{figure}[ht]
\begin{minipage}[t]{0.5\linewidth}
\centering
\includegraphics[width=\textwidth]{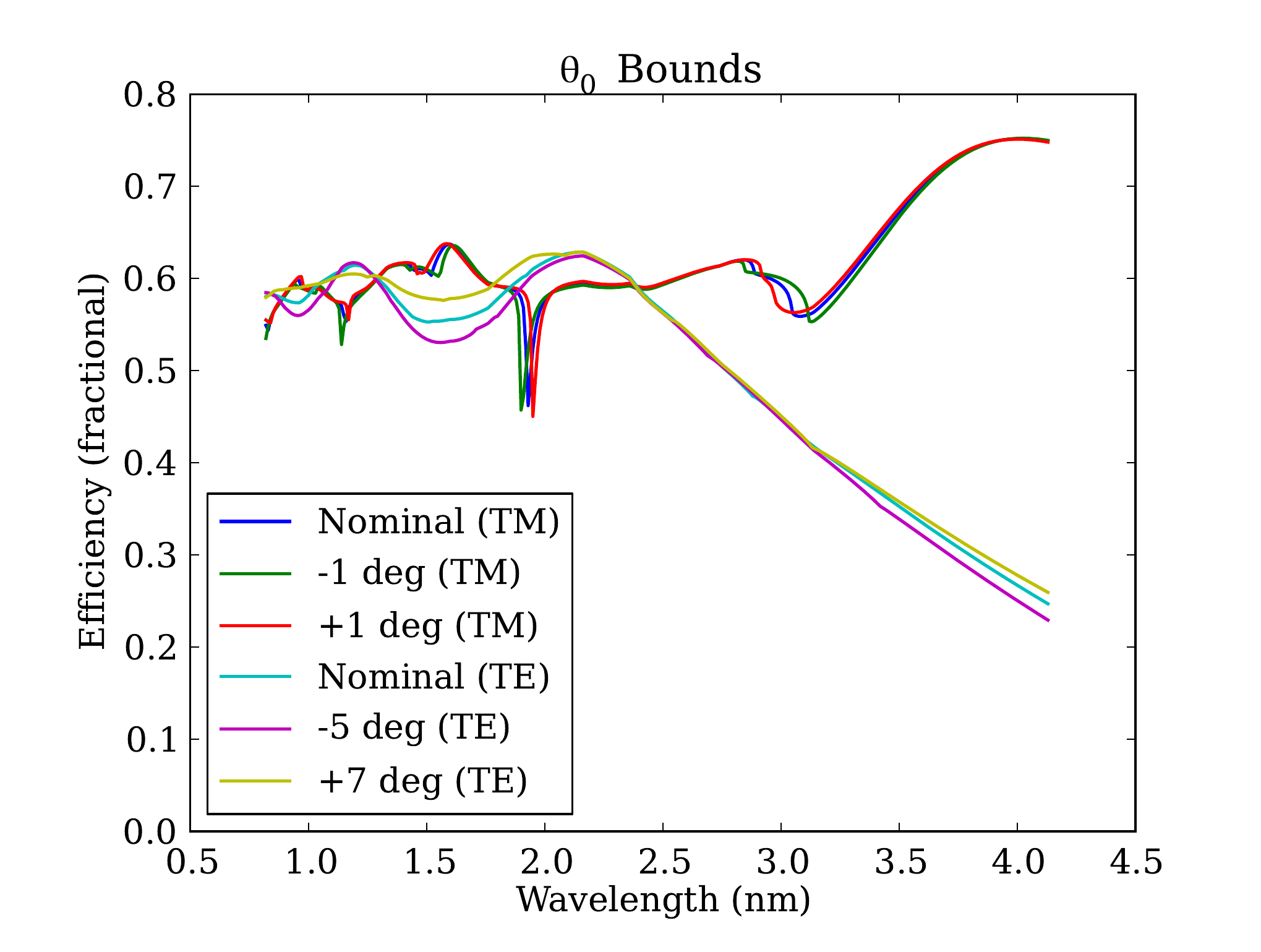}
\caption{Diffraction efficiencies for both transverse electric (TE) and transverse magnetic (TM) polarizations for a nominal ($18^\circ$) and misaligned $\theta_0$. For our angular misalignment tolerances, the maximum change in $\theta_0$ is $\pm22$ arcseconds (roll).}
\label{fig:ThetaBounds}
\end{minipage}
\hspace{0.5cm}
\begin{minipage}[t]{0.5\linewidth}
\centering
\includegraphics[width=\textwidth]{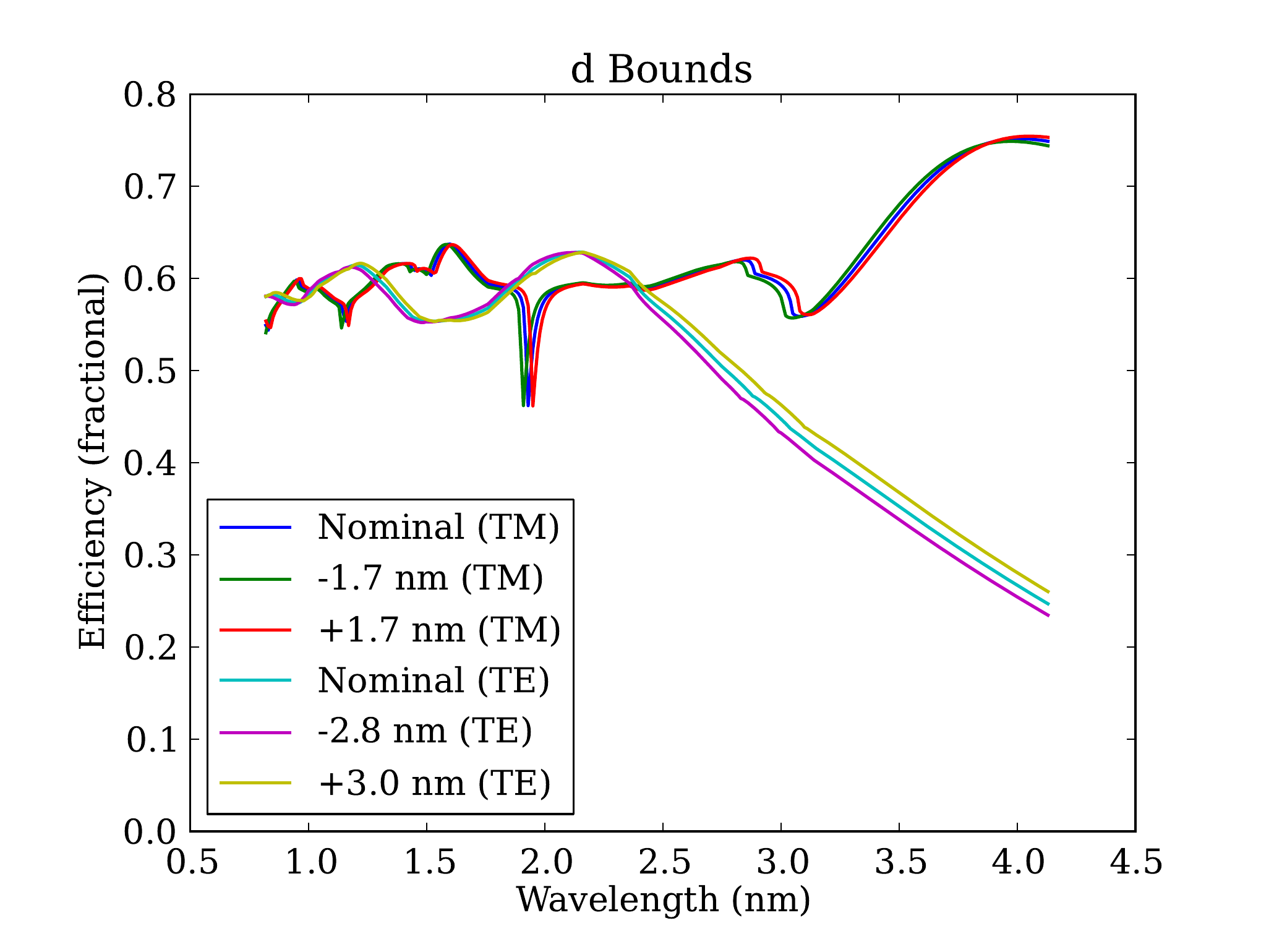}
\caption{Diffraction efficiencies for both transverse electric (TE) and transverse magnetic (TM) polarizations for a nominal (160 nm) and misaligned groove periods. For translational misalignments, the maximum change in groove period is $\pm0.03$ nm ($\hat z$ translation).}
\label{fig:DBounds}
\end{minipage}
\end{figure}

After an angular or translational misalignment, three parameters relating to diffraction efficiency can change: $\theta_0$, $\phi_0$, and groove period $d$.  Characteristic limits on these parameters are $\pm22$ arcseconds, $\pm4$ arcseconds, and $\pm0.03$ nm, obtained from the roll, pitch, and $\hat z$ translation tolerances above.  Our goal was to determine bounds on these parameters based on when the diffraction efficiency changes appreciably.  An appreciable change was defined as a $>1\%$ RMS difference in diffraction efficiency over the 300--1500 eV (4.1--0.8 nm) range.  For each parameter of interest, the value was shifted away from the nominal value in an iterative process until this change condition was reached.  This was done for both transverse electric (TE) and transverse magnetic (TM) polarizations.  Figs.\ \ref{fig:PhiBounds}, \ref{fig:ThetaBounds}, and \ref{fig:DBounds} show the nominal diffraction efficiency and the diffraction efficiency after the change condition was reached in both the positive and negative direction.  Note that the dramatic dependence on polarization is an expected result, and such a dependence has been measured in the past \cite{Seely}. In all cases, the misalignments which result in the 1\% RMS efficiency change are at least an order of magnitude greater than the characteristic limits given above.  They are also much greater than the pointing stability of a typical X-ray observatory ($\sim 0.25$ arcseconds for \textit{Chandra}).  These results indicate that a single diffraction efficiency curve can be used to determine effective area, and that the effective area (i.e.\ energy response function) can be assumed to be constant during a telescope pointing.

\section{Summary}

We have shown, using both an analytical and raytracing approach, the alignment tolerances for all six degrees of freedom for an off-plane reflection grating spectrometer.  We have used reasonable nominal alignment assumptions for a flight-like instrument, and have shown that our results scale linearly with focal length and spectral resolution and effective area requirements within realistic ranges.  Furthermore, we have shown that for our alignment tolerances, diffraction efficiency can be assumed to be constant.  In calculating the tolerance for a given misalignment, this work has assumed perfect alignment for the other five degrees of freedom.  In a later paper, we intend to show results of a raytracing algorithm designed to simultaneously incorporate all six alignment degrees of freedom.  We will present a best-case error budget for a flight-like spectrometer with a full grating array.

\begin{acknowledgements}
The authors acknowledge support from NASA Strategic Astrophysics Technology grant, NNX12AF23G.  The commercial software PCGrate-S(X) v.6.1 \copyright~(I.I.G.\ Inc.) was also crucial in the completion of this work.
\end{acknowledgements}



\end{document}